\documentclass[twocolumn,aps,prx,superscriptaddress]{revtex4-1}

\usepackage{graphicx}

\usepackage{multirow}

\begin{document}


\title{Uniaxially stressed germanium with fundamental direct band gap}


\author{R. Geiger}
\affiliation{Laboratory for Micro- and Nanotechnology, Paul Scherrer Institut, 5232 Villigen PSI, Switzerland}

\author{T. Zabel}
\affiliation{Laboratory for Micro- and Nanotechnology, Paul Scherrer Institut, 5232 Villigen PSI, Switzerland}

\author{E. Marin}
\affiliation{Laboratory for Micro- and Nanotechnology, Paul Scherrer Institut, 5232 Villigen PSI, Switzerland}

\author{A. Gassenq}
\affiliation{Universit{\'e} Grenoble Alpes, 38000, Grenoble, France}
\affiliation{CEA-INAC, SP2M Minatec Campus, 17 Rue de Martyrs, 38000, Grenoble, France}

\author{J.-M. Hartmann}
\affiliation{Universit{\'e} Grenoble Alpes, 38000, Grenoble, France}
\affiliation{CEA-LETI, Minatec Campus, 17 Rue de Martyrs, 38054, Grenoble, France}

\author{J. Widiez}
\affiliation{Universit{\'e} Grenoble Alpes, 38000, Grenoble, France}
\affiliation{CEA-LETI, Minatec Campus, 17 Rue de Martyrs, 38054, Grenoble, France}

\author{J. Escalante}
\affiliation{Universit{\'e} Grenoble Alpes, 38000, Grenoble, France}
\affiliation{CEA-INAC, SP2M Minatec Campus, 17 Rue de Martyrs, 38000, Grenoble, France}

\author{K. Guilloy}
\affiliation{Universit{\'e} Grenoble Alpes, 38000, Grenoble, France}
\affiliation{CEA-INAC, SP2M Minatec Campus, 17 Rue de Martyrs, 38000, Grenoble, France}

\author{N. Pauc}
\affiliation{Universit{\'e} Grenoble Alpes, 38000, Grenoble, France}
\affiliation{CEA-INAC, SP2M Minatec Campus, 17 Rue de Martyrs, 38000, Grenoble, France}

\author{D. Rouchon}
\affiliation{Universit{\'e} Grenoble Alpes, 38000, Grenoble, France}
\affiliation{CEA-LETI, Minatec Campus, 17 Rue de Martyrs, 38054, Grenoble, France}

\author{G. Osvaldo Diaz}
\affiliation{Universit{\'e} Grenoble Alpes, 38000, Grenoble, France}
\affiliation{CEA-LETI, Minatec Campus, 17 Rue de Martyrs, 38054, Grenoble, France}

\author{S. Tardif}
\affiliation{Universit{\'e} Grenoble Alpes, 38000, Grenoble, France}
\affiliation{CEA-INAC, SP2M Minatec Campus, 17 Rue de Martyrs, 38000, Grenoble, France}

\author{F. Rieutord}
\affiliation{Universit{\'e} Grenoble Alpes, 38000, Grenoble, France}
\affiliation{CEA-INAC, SP2M Minatec Campus, 17 Rue de Martyrs, 38000, Grenoble, France}

\author{I. Duchemin}
\affiliation{Universit{\'e} Grenoble Alpes, 38000, Grenoble, France}
\affiliation{CEA-INAC, SP2M Minatec Campus, 17 Rue de Martyrs, 38000, Grenoble, France}

\author{Y.-M. Niquet}
\affiliation{Universit{\'e} Grenoble Alpes, 38000, Grenoble, France}
\affiliation{CEA-INAC, SP2M Minatec Campus, 17 Rue de Martyrs, 38000, Grenoble, France}

\author{V. Reboud}
\affiliation{Universit{\'e} Grenoble Alpes, 38000, Grenoble, France}
\affiliation{CEA-LETI, Minatec Campus, 17 Rue de Martyrs, 38054, Grenoble, France}

\author{V. Calvo}
\affiliation{Universit{\'e} Grenoble Alpes, 38000, Grenoble, France}
\affiliation{CEA-INAC, SP2M Minatec Campus, 17 Rue de Martyrs, 38000, Grenoble, France}

\author{A. Chelnokov}
\affiliation{Universit{\'e} Grenoble Alpes, 38000, Grenoble, France}
\affiliation{CEA-LETI, Minatec Campus, 17 Rue de Martyrs, 38054, Grenoble, France}

\author{J. Faist}
\affiliation{Institute for Quantum Electronics, ETH Z{\"u}rich, 8093 Z{\"u}rich, Switzerland}

\author{H. Sigg}
\email[]{hans.sigg@psi.ch}
\affiliation{Laboratory for Micro- and Nanotechnology, Paul Scherrer Institut, 5232 Villigen PSI, Switzerland}


\begin{abstract}
We demonstrate the crossover from indirect- to direct band gap in tensile-strained germanium by temperature-dependent photoluminescence. The samples are strained microbridges that enhance a biaxial strain of 0.16\% up to 3.6\% uniaxial tensile strain. Cooling the bridges to 20 K increases the uniaxial strain up to a maximum of 5.4\%. Temperature-dependent photoluminescence reveals the crossover to a fundamental direct band gap to occur between 4.0\% and 4.5\%. Our data are in good agreement with new theoretical computations that predict a strong bowing of the band parameters with strain.
\end{abstract}

\pacs{}

\maketitle


On-chip data transmission is currently the bottleneck to further increase computing power as metal wiring reaches its fundamental limits concerning band width and energy consumption \cite{Miller:2009fi}. Hence, optical interconnects are envisioned to overcome the drawbacks of their electrical counterpart. However, the realization of an efficient light source compatible with the complementary metal-oxide-semiconductor (CMOS) environment is still the greatest challenge for the convergence of electronics and photonics on silicon. This is caused by the indirect nature of silicon`s band gap which prohibits efficient light emission. For monolithic integration of a light source on Si it is, hence, desirable to have a material with a direct band gap but - in contrast to direct gap III-V lasers heterogeneously integrated on Si \cite{Fang:2008kr} - with chemical compatibility to Si. In this respect, Ge has gained a lot of attention due to its CMOS compatibility and its small conduction band offset between the direct $\Gamma$ and the indirect L states of only $\sim$ 140 meV. To close this offset and transform Ge into a direct band gap semiconductor, the application of tensile strain \cite{Suess:2013ks, Ghrib:2015jb, Capellini:2014ec} as well as alloying Ge with Sn \cite{Wirths:2015dh, Chen:2013fy, Gallagher:2015kb} have become a very active field of research. The recent observation of lasing in a high Sn-content partially strain-relaxed GeSn-alloy with a band offset of minus 25 meV (i.e. minimum of $\Gamma$-valley below the energy of the L-valleys) delivered a proof of concept for direct band gap group IV light emitters \cite{Wirths:2015dh}. However, lasing was so far limited to a maximum temperature of 90 K due to the rapidly decreasing non-radiative lifetime of only a few hundred picoseconds at temperatures $>$ 100 K \cite{Wirths:2015dh}. Direct band gap elemental Ge of high quality would, in contrast, take full advantage from longer non-radiative recombination times enabling lasing at elevated temperatures given that a similar conduction band offset can be achieved. However, in spite of the many efforts to reach a direct band gap configuration in Ge by applying tensile strain, there is still the lack of proving a fundamental direct band gap as spectroscopy on structures with a sufficiently high strain was either not presented \cite{Sukhdeo:2014dd} or performed only at room temperature \cite{Ghrib:2015jb, SanchezPerez:2011vq}, which is not sufficient to substantiate the claim.\par
 
In this letter, we present Ge microstructures fabricated from high-quality optical germanium-on-insulator (GeOI) substrates where the material quality is assessed by determination of the minority carrer lifetime. By measuring temperature-dependent photoluminescence (PL) on structures with different strain and at various excitation intensities, we validate the transition from an indirect- to a fundamental direct band gap semiconductor - which manifests in a strong direct gap emission at 20 K - to occur for Ge with a uniaxial strain between 4.0\% and 4.5\%.\par 

 \begin{figure*}
 	\includegraphics[width=\textwidth]{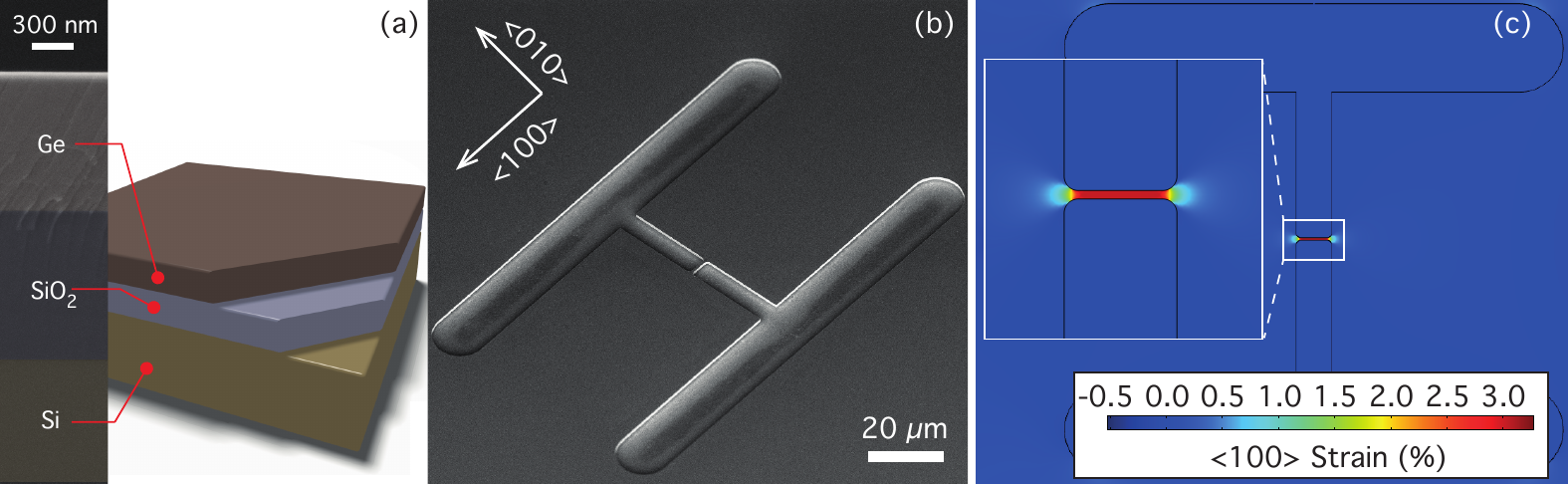}%
 	\caption{(a) Schematics and cross-sectional scanning electron microscope (SEM) image of an optical germanium-on-insulator (GeOI) substrate. (b) SEM top-view of a suspended GeOI microbridge. (c) Finite-element-method simulation of $<$100$>$ strain distribution in a GeOI microbridge.}
 \end{figure*}
 
To introduce a high tensile stress in Ge, we make use of stress-enhancement by geometrical patterning as introduced in ref. \cite{Suess:2013ks}. GeOI serves as the starting substrate. It is obtained from a thick layer of intrinsic Ge grown on Si which gets transferred to an oxidized Si wafer using SmartCut$^{\mathrm{TM}}$ technology \cite{Reboud:2015hw}. After layer transfer, the misfit dislocations originating from the defective Ge/Si interface region are located on the top of the layer stack and are, therefore, readily accessible and removed via chemical mechanical polishing (CMP). After CMP, a 1 $\mu$m thick Ge layer of high quality is obtained on top of a 1 $\mu$m buried oxide layer (see Fig. 1(a) and supporting material (SM) \cite{Anonymous:cZhqGTFU}). The bonding process preserves most of the epitaxially induced strain which originates from the difference in thermal expansion coefficients between Si and Ge, leaving a biaxial tensile strain of 0.16\% at room temperature after layer transfer and CMP.\par
Patterning of the Ge layer into $<$100$>$ oriented microbridges is performed via electron beam lithography and reactive ion etching to define a central, narrow ‘constriction’ symmetrically surrounded by ‘pads’ with large cross sections \cite{Suess:2013ks}, see Fig. 1(b). When the underlying buried oxide is selectively removed with vaporous hydrofluoric acid, the pads relax which leads to a strong uniaxial tensile stress in the constriction, as accurately reproduced by finite-element COMSOL (FEM) modelling shown in Fig. 1(c). This geometrical enhancement of tensile strain can be widely tuned by choosing appropriate ratios of width and length for constriction and pads with the material`s yield strength as the limiting parameter \cite{Suess:2013ks}.\par

  \begin{figure*}
  	\includegraphics[width=\textwidth]{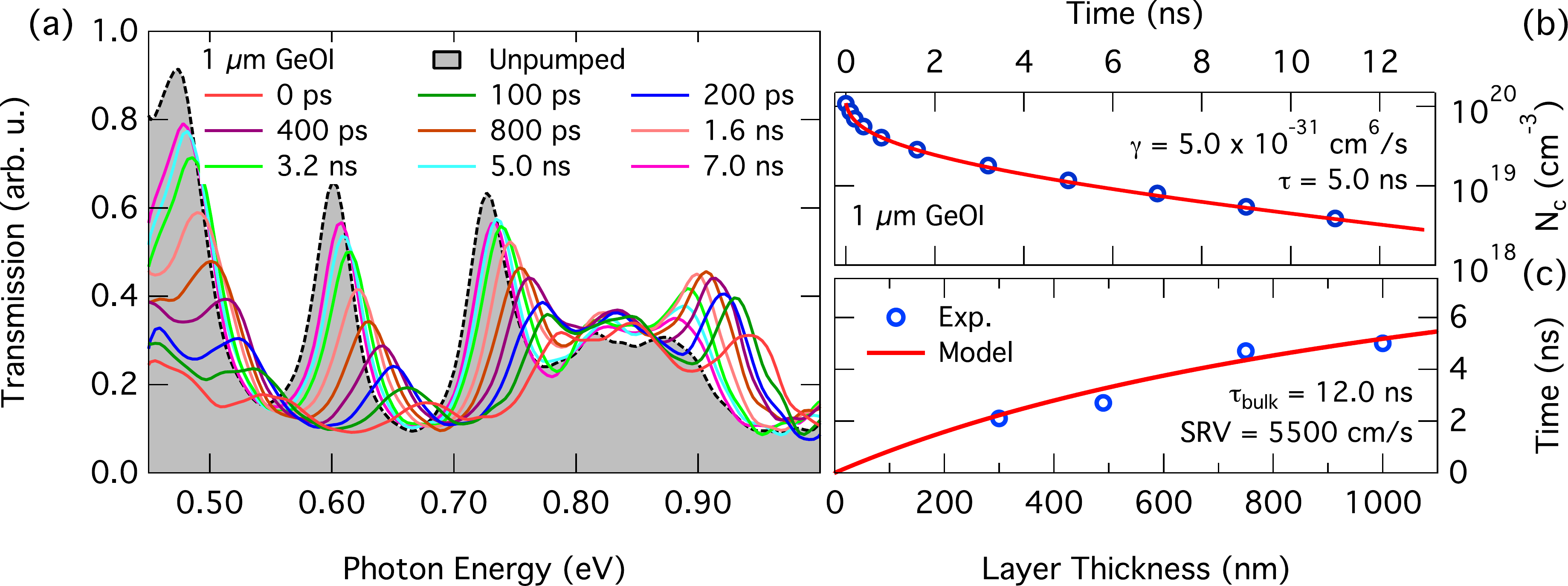}%
  	\caption{(a) Broadband pump-probe transmission spectra for a 1 $\mu$m thick GeOI epilayer at varying pump-probe delay times. (b) Charge carrier densities $N_c$ extracted from the spectra shown in (a). A lifetime of 5.0 ns is extracted from an exponential fit to the decay as well as an Auger coefficient of 5.0$\times$10$^{-31}$ cm$^6$/s. (c) Decay time as a function of layer thickness. A bulk lifetime of 12.0 ns is extracted from the experimental data.}
  \end{figure*}

The motivation to use GeOI as starting substrate lies in the improved mechanical and optical properties compared to Ge layers directly on Si: Due to the reduced density of threading dislocations and the absence of a defective Ge/Si interface, the yield strength increases and the internal quantum efficiency of light emission rises thanks to a longer non-radiative recombination time. We extract the latter via infrared pump-probe transmission measurements at the infrared beamline X01DC of the Swiss Light Source using a 100 ps Nd:YAG laser pulse as excitation and the broadband synchrotron radiation as probe \cite{Carroll:2011vi}. In Fig. 2(a), transmission spectra through a GeOI substrate are shown for varying delay times between pump- and probe pulses. The high refractive index contrast between Ge and the buried oxide facilitates distinct Fabry-Perot (FP) interferences. The peak positions shift with a linear dependence on the optically excited charge carrier density $N_c$ as $\Delta E/E$ is proportional to $N_c/n_r^2 E^2$, where $n_r$ is the refractive index \cite{Geiger:2014gw}. From the spectra shown in Fig. 2(a), the extracted charge carrier densities (blue circles) are plotted together with the fitted exponential decay (red line) in Fig. 2(b). The non-radiative lifetime $\tau$ as well as the Auger recombination coefficient $\gamma$, which relates the Auger recombination time $\tau_A$ to the carrier density as $\tau_A = 1/\gamma N_c^2$, serve as free fitting parameters, which are obtained as $\tau$ = 5.0 ns and $\gamma$ = 5.0$\times$10$^{-31}$ cm$^6$/s (for details see \cite{Anonymous:cZhqGTFU}). The latter value is a factor 2 to 3 larger than compared to literature values obtained under low excitation \cite{Uleckas:2011ds} but is about one order of magnitude lower than the Auger recombination rate as obtained under carrier saturation conditions \cite{Carroll:2012hw} when probably also higher than second order Auger processes begin to play a role. To decouple bulk- and surface/interface effects, the lifetime is measured on samples with different thickness after thinning by reactive ion etching using SF$_6$, Ar and CHF$_3$ (see Fig. 2(c)). The thickness-dependent data are accurately described by

\begin{equation}
\frac{1}{\tau} = \frac{1}{\tau_B} + \left( \frac{d}{2S} + \frac{d^2}{\pi^2 D} \right)^{-1},
\end{equation}
\\
where $\tau_B$ is the bulk lifetime, $d$ the layer thickness, $S$ the surface recombination velocity and $D$ the diffusion constant \cite{Saito:2014gs, Gaubas:2006eo, Sproul:1994dz}. With $D$ = 100 cm$^2$/s \cite{Young:1982hf}, we obtain $S$ = 5500 cm/s and a bulk lifetime of $\tau_B$ = 12.0 ns. Apparently, the dry etching does not affect the surface recombination velocity which we attribute to a passivation of the surface states with methyl groups \cite{Jiang:2014bn}.\par

 \begin{figure*}
 	\includegraphics[height=6.9cm]{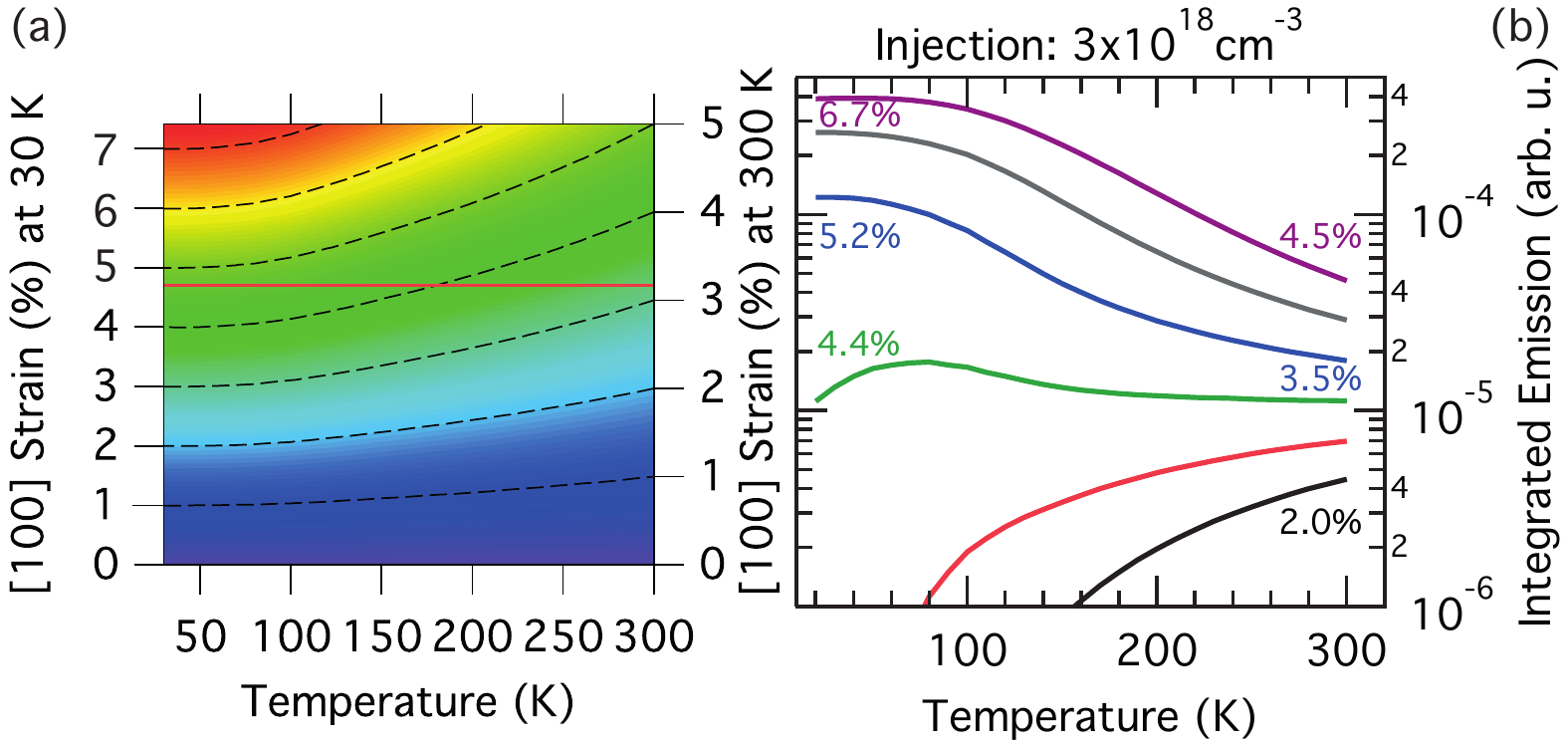}%
 	\caption{(a) Contour-plot of calculated uniaxial tensile strain in a microbridge. Due to the difference in thermal expansion coefficient between Si and Ge, the strain increases as the temperature is decreased. The red line shows that 3.2\% strain at room temperature are sufficient to achieve 4.7\% at 50 K. (b) Model for integrated direct gap emission as a function of temperature for a constant injected carrier density of 3$\times$10$^{18}$ cm$^{-3}$. The numbers indicate the strain at 300 K or 20 K, respectively.}
 \end{figure*}

Using the unthinned GeOI material, microbridges with constrictions of 6 $\mu$m $\times$ 500 nm and varying pad lengths are fabricated, yielding a maximum Raman shift of -6.7 cm$^{-1}$ from power-dependent Raman spectroscopy \cite{Suess:2014by}. According to the recently determined non-linear relation between Raman shift $\Delta \omega$ and  $<$100$>$ uniaxial strain ε given in ref. \cite{Gassenq:VgOg3Rm6}, this corresponds to a strain of 3.6\% at room temperature. By cooling the samples, this strain is enhanced due to the different thermal expansion coefficients of Si and Ge and the redistribution of the tensile strain in the Ge layer by $\varepsilon_L \times EF$, where $EF$ is the enhancement factor given by the structure`s geometry and $\varepsilon_L$ denotes the biaxial strain in the layer. Taking into account the temperature-dependent thermal expansion coefficients of Si 
\cite{Okada:1984tl} and Ge \cite{Novikova:1960vf}, the biaxial strain in the layer increases by 0.073\% to a total of $\sim$ 0.24\% when going from 300 K down to 20 K. As shown in Fig. 3(a), the strain in a given microbridge increases accordingly. From the finite element modelling, it follows that e.g. 4.7\% - which, according to deformation potential theory, appears to be the crossover to a fundamental direct band gap \cite{Geiger:2015jp} - can already be obtained for bridge structures with a room-temperature strain of 3.2\% when cooling to temperatures below 50 K as highlighted in Fig. 3(a).\par

  \begin{figure*}[t!]
  	\includegraphics[width=\textwidth]{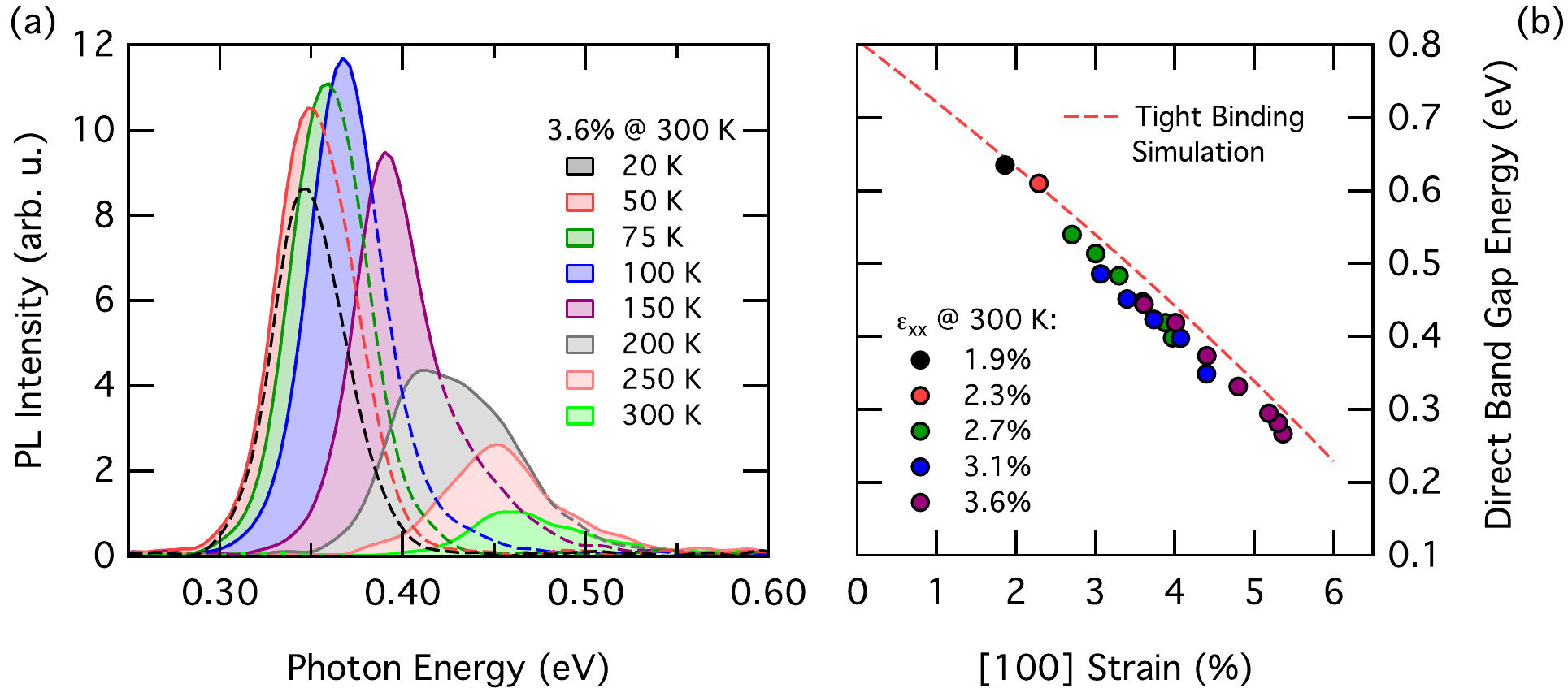}%
  	\caption{(a) PL spectra at varying temperatures for a structure with 3.6\% uniaxial tensile strain at 300 K. The T-dependent strain increase leads to a red-shift in emission energy. (b) Direct gap energies extracted from temperature-dependent PL measurements on different microbridges and shifted to their respective value at 300 K via Varshni`s law. The strain values are given at 300 K. The red, broken line is calculated via tight-binding modelling.}
  \end{figure*}
  
Besides the increase in strain, lowering the temperature also thermalizes the electrons into the lowest energy states available in the conduction band. This effect will be exploited here to distinguish between fundamentally direct- and indirect band gap, as we have recently done for the proof of a fundamental direct band gap in GeSn \cite{Wirths:2015dh}. In Fig. 3(b), the integrated spontaneous emission intensity for direct gap transitions is calculated as a function of temperature for structures with room-temperature (RT) strain values between 2.0\% and 4.5\% at a constant injected carrier density of $3\times 10^{18}$ cm$^{-3}$. The model takes into account the T-dependent strain increase from Fig. 3(a) with linear band shifts according to deformation potential theory \cite{Geiger:2015jp}, as well as the joint-density-of-states of the dipole-allowed transitions between the $\Gamma$-valley and the highest valence band states, assuming isotropic, parabolic bands (see ref. \cite{Wirths:2015dh} for details). Two regimes are found: (i) For $\varepsilon <$ 3.0\% at RT, the direct gap – crossover cannot be reached and the electrons, hence, populate the indirect L-valleys, which does not yield efficient radiative recombination. (ii) For $\varepsilon >$ 3.0\% at RT, however, cooling increases the population of the $\Gamma$-valley such that the direct gap emission intensity increases.\par

In Fig. 4(a), PL spectra of a microbridge with 3.6\% strain at 300 K are shown for varying temperatures. Carriers are optically excited by a continuous-wave laser emitting at 532 nm with an incident power of 7 mW on an approximately 7 $\mu$m spot corresponding to $\sim$ 18 kW/cm$^2$. As expected from our model in Fig. 3(a), the strain increases for lower temperatures, leading to a red-shift in emission which overcompensates the typically observed blue-shift of the band gap. To further corroborate the validity of our temperature-dependent strain extrapolation, we extract the strain-dependent direct band gaps $E_{gap}$ for a set of samples with different strain by fitting the PL peak with a simple model for the spontaneous radiative efficiency $R(E)$ of a bulk material as

\begin{figure*}[t!]
	\includegraphics[width=\textwidth]{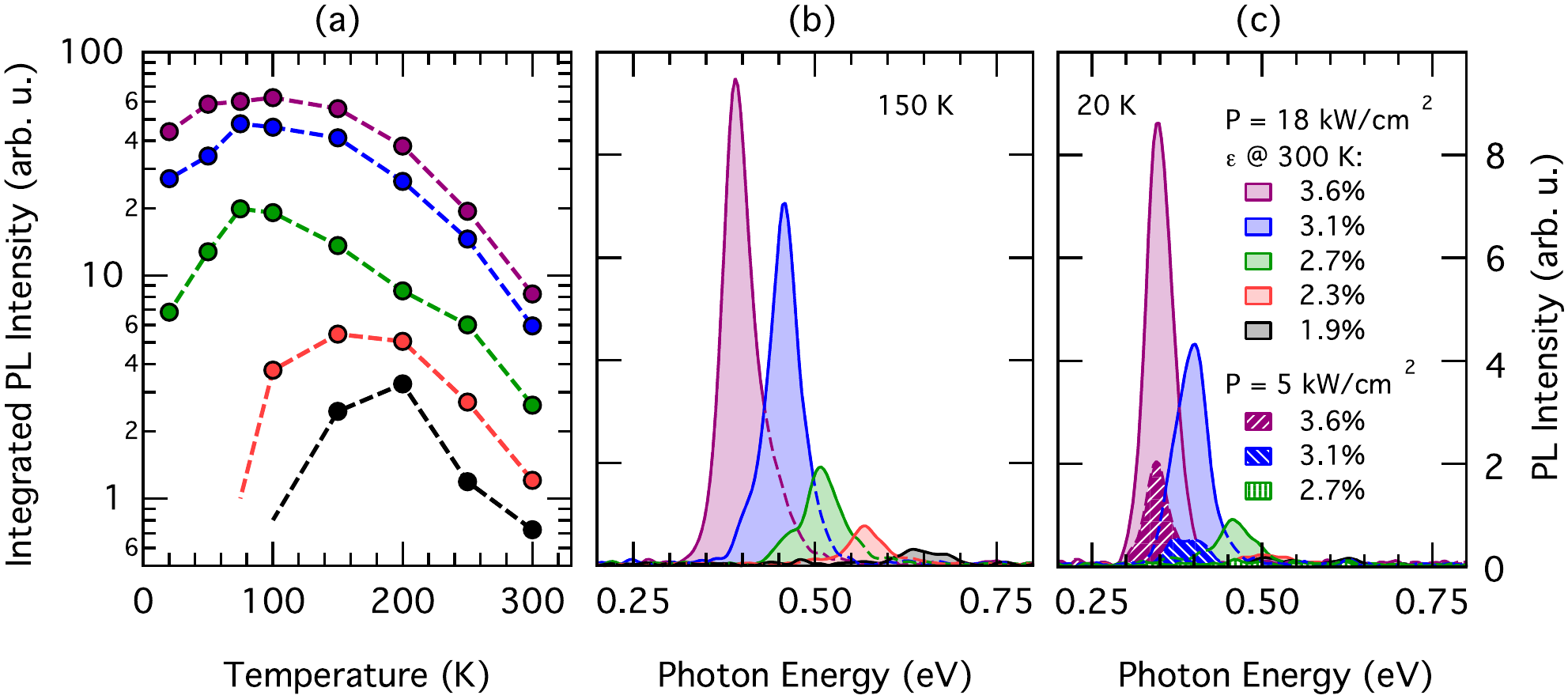}%
	\caption{(a) Integrated direct gap emission intensities as a function of temperature for a set of differently strained microbridges. Please note that the ordinate scale is logarithmic, i.e. that intensity increases are substantial. (b) PL spectra at 150 K as a function of uniaxial tensile strain. (c) PL spectra at 20 K for excitation densities of 18 kW/cm$^2$ (filled, solid curves) and 5 kW/cm$^2$ (filled, shaded curves).}
\end{figure*}

\begin{equation}
	R\left(E\right) = \sqrt{E-E_{gap}} \exp{\left(\frac{E}{kT}\right)},
\end{equation}
\\
where $k$ is the Boltzmann constant. The obtained band gaps are shifted towards their room-temperature value using Varshni`s formula \cite{Varshni:1967gl}:

\begin{equation}
	E_{gap} \left( T \right) = E_{gap} \left( 0 \right) - \frac{a \, T^2}{T+b}.
\end{equation}
\\
Here, $E_{gap}\left(0\right)$ = 0.88 eV (ref. \cite{Schaffler:bb}), $a$ = 4.774$\times$×10$^{-4}$ 1/K (ref. \cite{Thurmond:1975wv}) and b = 235 K (ref. \cite{Thurmond:1975wv}). The extrapolation of strain values for low temperatures according to FEM simulations seems confirmed as the band gaps extracted from different structures (i.e. with different RT strain) at the same strain state show an excellent agreement (see Fig. 4(b)). It should be noted that even after several cycles of cooling, the structures show no sign of degradation and the same maximum strain is reached. In Fig. 4(b), the experimental band gap energies at $\Gamma$ are compared to those calculated with the tight-binding model and methodology of ref. \cite{Niquet:2009ju} which was previously applied for biaxially strained Ge. This model reproduces \textit{ab initio} calculations over a wide range of deformations and predicts, in particular, a bowing of band gap energy at large uniaxial strain which is found to be in good agreement with the experimental data.\par  
Lastly, the intensities integrated over the PL peak are plotted in Fig. 5(a) as a function of temperature for the excitation density of 18 kW/cm$^2$. When comparing intensities between the different samples at any given temperature, higher strain yields higher emission intensity due to a progressively more favorable alignment between $\Gamma$- and L-valleys. When cooling from 300 K down to 200 K, we observe an intensity increase for all bridges including the ones at strain below 3.0\% albeit a monotonic decrease would be expected according to Fig. 3(b). We attribute this to the ambipolar carrier diffusion which helps to collect a large amount of excited carriers from the pad into the bridge region, in particular at lower temperatures (we estimate a $\sim10\times$ larger carrier density to be excited outside of the constriction compared to the highly strained part, see \cite{Anonymous:cZhqGTFU}). This experimentally observed effect \cite{Suess:2013ks,Nam:2013gg} is not accounted for in the model. In contrast, the faster carrier diffusion will lead to an enhanced surface recombination velocity \cite{Stevenson:1954js}. However, this effect seems to be overcompensated by the carrier collection. At 150 K, direct gap emission from all samples can still be detected (see Fig. 5(b)), whereas for lower temperatures the emission for bridges with $\varepsilon <$ 2.7\% at RT vanishes.\par

At 20 K, a strong direct gap emission is obtained from samples with RT strain of 2.7\%, 3.1\% and 3.6\% translating into 4.0\%, 4.5\% and 5.4\% at 20 K (Fig. 5(c)). Under a lower excitation density of 5 kW/cm$^2$ which corresponds to a steady-state carrier density of 4.6$\times$10$^{17}$ cm$^{-3}$ at a lifetime of 5.0 ns and neglecting carrier diffusion, the direct gap emission of the 4.5\% and 5.4\% strained bridges can still be clearly detected, while the intensity at 4.0\% strain drops below the noise limit (see the shaded, filled areas in Fig. 5(c)). Therefore, samples at 4.5\% and 5.4\% are identified as having a fundamental direct band gap, whereas 4.0\% appears to be around the transition point. The crossover of Ge towards a true direct band gap semiconductor is, hence, found between 4.0\% and 4.5\% uniaxial strain, which is the main outcome of this letter.\par 
The intensity decrease for direct gap samples when cooling from 100 K to 20 K cannot be explained by above introduced model that predicts a constant PL intensity as a function of temperature in that temperature range. We ascribe this to a lower efficiency for phonon-assisted L$\rightarrow \Gamma$ intervalley scattering (IVS) for electrons diffusing from the outer pad regions (where the population is located in the indirect valley) into the central constriction (with its direct band gap), similar to experimental results for temperature-dependent X$\rightarrow \Gamma$,L scattering times in GaAs \cite{Cavicchia:1996ik}. Due to the near-degeneracy of the $\Gamma$- and L-valleys in the strained Ge, the IVS phonon bottleneck is more pronounced than in GaAs, which explains that in spite of the long non-radiative lifetime, the occupation of $\Gamma$ at low temperature does not seem to be in thermal equilibrium with the one of the L valley. This phonon bottleneck will disappear when the $\Gamma$ band edge is more than a typical phonon energy (30 meV) below the L states. This is exactly what we observe for the sample with the highest strain where we extrapolate that the band gap offset is approximately minus 40 meV with the decline being less than a factor of 2. For this extrapolation, we assume the crossover at 4.25 \% and use a simple linear correlation, i.e. the offset value is 140 meV $\times$ $\left( 1-5.40/4.25\right)$ = -40 meV. An alternative explanation would attribute the behavior to a different coefficient in Varshni`s law for $\Gamma$ and L states \cite{Varshni:1967gl}, as a weaker T-dependent energy decrease for L than for $\Gamma$ would have a similar effect. However, as the band gap variations vanish towards low temperatures, the Varshni parameter-argument seems less likely for a sound explanation of the observed effect.\par
In summary, we have investigated strained germanium microbridges fabricated from high-quality germanium-on-insulator substrates with strain values around the transition from indirect to fundamental direct band gap. Temperature-dependent photoluminescence reveals a qualitatively different behavior depending on strain, with a vanishing direct gap emission at low temperature for $\varepsilon <$ 4.0\% and a strong direct gap emission for higher strain. Therefore, the crossover of germanium towards an elemental group IV semiconductor with a fundamental direct band gap was evidenced at a strain between 4.0\% and 4.5\%. Furthermore, we have shown long carrier lifetimes in optical GeOI substrates and observed a phonon bottleneck for thermalization of the carriers from the L- into the $\Gamma$ states. Finally, our data are in agreement with a strong bowing of the band gap energy with strain as predicted by theory. In conclusion, long non-radiative lifetimes together with high mechanical stability make the newly developed GeOI substrate a most promising platform to realize an efficient direct band gap laser for monolithic integration on a Si CMOS platform.

\begin{acknowledgments}
Parts of this work were supported by the Swiss National Science Foundation (SNF Project No. 149294) as well as by the CEA projects DSM-DRT Phare Photonics and Operando. We thank Stefan Stutz for sample preparation and the X01DC beamline of the Swiss Light Source for accessibility.
\end{acknowledgments}

\bibliography{parts/Papers_References_All}


	%
	%
	\pagebreak
	\widetext
	\begin{center}
		\textbf{\large Supplemental Material for: Uniaxially stressed germanium with fundamental direct band gap}
	\end{center}
	
	\setcounter{equation}{0}
	\setcounter{figure}{0}
	\setcounter{table}{0}
	\setcounter{page}{1}
	\makeatletter
	\renewcommand{\theequation}{SM\arabic{equation}}
	\renewcommand{\thefigure}{SM\arabic{figure}}
	\renewcommand{\thetable}{SM\Roman{table}}
	\renewcommand{\bibnumfmt}[1]{[SM#1]}
	\renewcommand{\citenumfont}[1]{SM#1}

		\section{Germanium-on-insulator fabrication process}
		
		The process steps for the fabrication of germanium-on-insulator (GeOI) are schematically shown in Fig. SM1. At first, 2.5 $\mu$m Ge is grown on 200 mm Si wafers via reduced-pressure chemical vapor deposition, followed by 200 nm SiO$_2$ via plasma-enhanced chemical vapor deposition. Subsequently, a defect layer is locally created in the Ge layer by H$^+$ ion implantation. Then, the wafer is bonded onto a silicon wafer which is covered with a thick thermal SiO$_2$ layer. During the anneal of the bonded wafer stack, the pair splits at the defect layer created by ion implantation. Finally, the surface of the Ge layer is treated by chemical mechanical polishing.
		
		\begin{figure}[h]
			\includegraphics{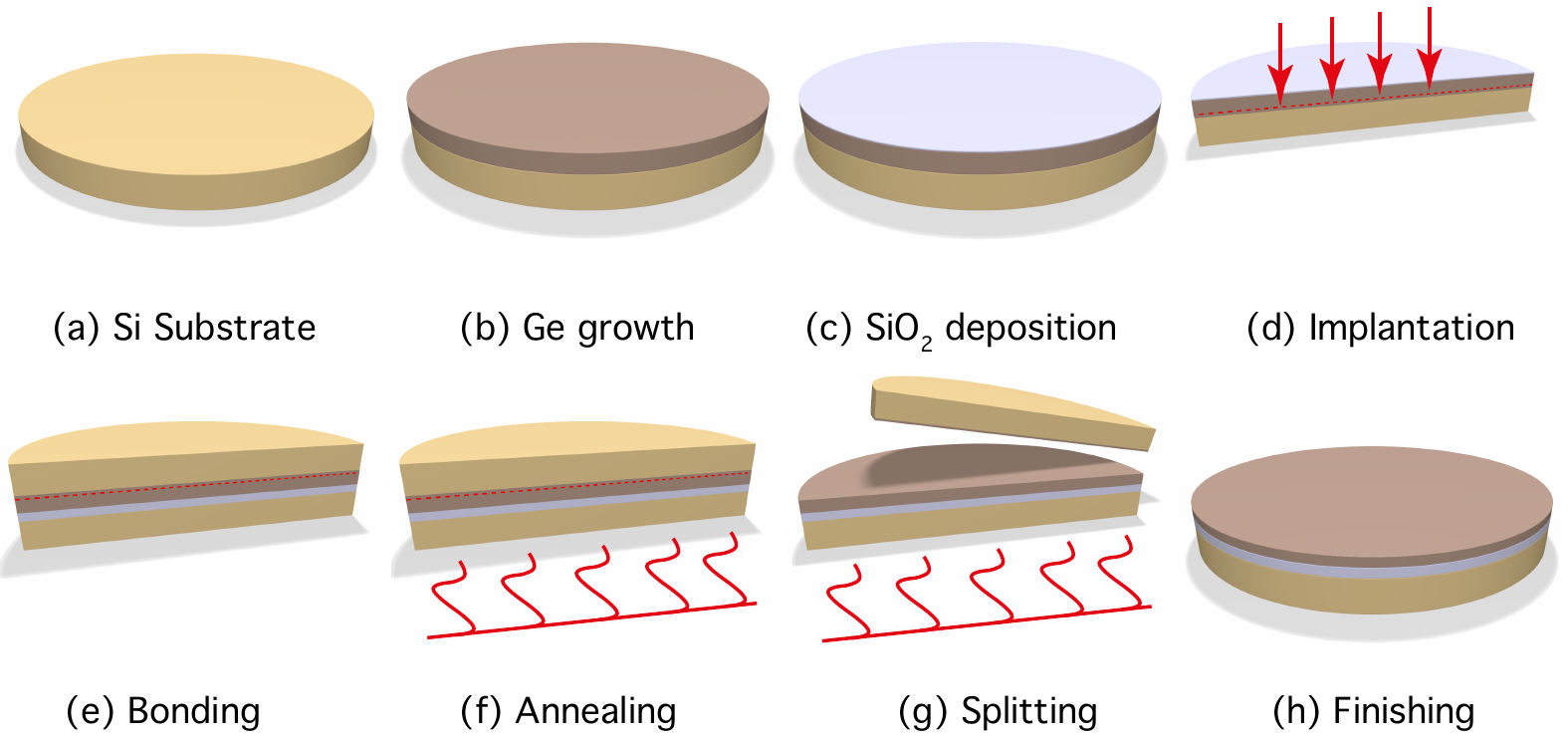}%
			\caption{Process flow for the fabrication of high-quality optical germanium-on-insulator substrates using the SmartCut$^\mathrm{TM}$ technology.}
		\end{figure}
		
		
		\section{Modelling of carrier decay}
		
		From time-resolved synchrotron-based pump-probe transmission measurements, we extract the charge carrier density and fit the decay with a simple model including a non-radiative decay time $\tau$ and a decay time $\tau_A$ related to Auger processes. The decay of the charge carrier density $N_c$ is described by the equation
		
		\begin{equation}
		N_c \left( t \right) = N_0 \left( \exp\left( -\frac{t}{\tau_A} \right) + \exp\left( -\frac{t}{\tau} \right)  \right),
		\end{equation}
		\\
		where $\tau_A$ is expressed with the Auger recombination coefficient $\gamma$ as $\tau_A$ = 1/$\gamma N_c^2$. After solving equation (1) for $N_c\left(t\right)$ and fitting the model to the experimental data, we obtain the non-radiative recombination time $\tau$ and the Auger coefficient $\gamma$ as 5.0 ns and 5.0$\times$10$^{-31}$ cm$^6$/s, respectively. The robustness of the fit towards changes in $\gamma$ is shown in Fig. SM2, where the decay model is plotted for three different Auger coefficients $\gamma$, revealing that the coefficient can be determined with an accuracy of $<$ 1.0$\times$10$^{-31}$ cm$^6$/s.

		\begin{figure}[h]
			\includegraphics[height=7.9cm]{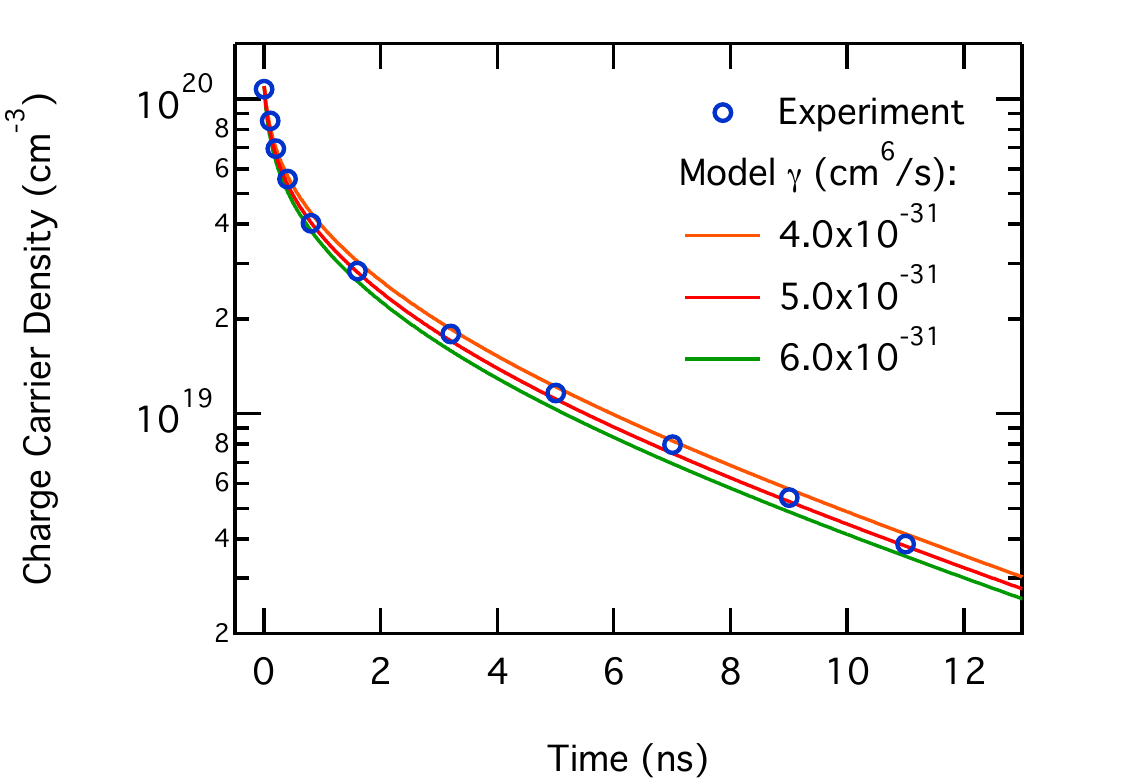}%
			\caption{Experimentally determined charge carrier densities in function of time. To visualize the accuracy for the extracted Auger recombination rate $\gamma$, three calculated decay curves are shown for Auger rates of 4$\times$10$^{-31}$, 5$\times$10$^{-31}$ and 6$\times$10$^{-31}$ cm$^6$/s, respectively.}
		\end{figure}
		
		\newpage
		
		\section{Estimation of carrier diffusion into constriction center}
		
		For temperature-dependent photoluminescence measurements, it is found that the luminescence increases for all structures upon cooling, although the calculated carrier population at the direct $\Gamma$ valley for the samples with indirect band gap (strain at 300 K smaller than approx. 3\%, c.f. Fig. 3(b) in main text) decreases. We attribute this to the fact that the model does not take into account the diffusion of carriers from regions in close vicinity of the highly strained constriction into the center of the structure, which was previously shown to contribute to the intensity of the emission at maximum strain [SM1]. The same carrier collection effect has been acknowledged by Nam \textit{et al.} from exploring so-called strain-induced pseudo-heterostructure nanowires [SM2]. \par
		For the here investigated structures, the area with a homogeneous distribution at maximum strain level is limited to $\sim$ 4.5 $\mu$m $\times$ 500 nm (c.f. finite element COMSOL modelling shown in Fig. SM3(a)). The excitation spot, however, extends over a larger area and is estimated to a width of 2$\sigma$ $\sim$ 7 $\mu$m which means that most of the carriers are excited outside of the region with the maximum strain. This is shown in Fig. SM3(b) where the area of excitation weighted with a 2-dimensional symmetric Gaussian distribution is compared to the area of the homogeneously strained part of the constriction. Depending on the collection efficiency, which itself depends on the temperature due to the increase in mobility, additional carriers will flow into the central region of the bridge where they recombine and thus enhance the PL efficiency, as is observed in the experiment.

		\begin{figure}[t!]
			\includegraphics[width=16cm]{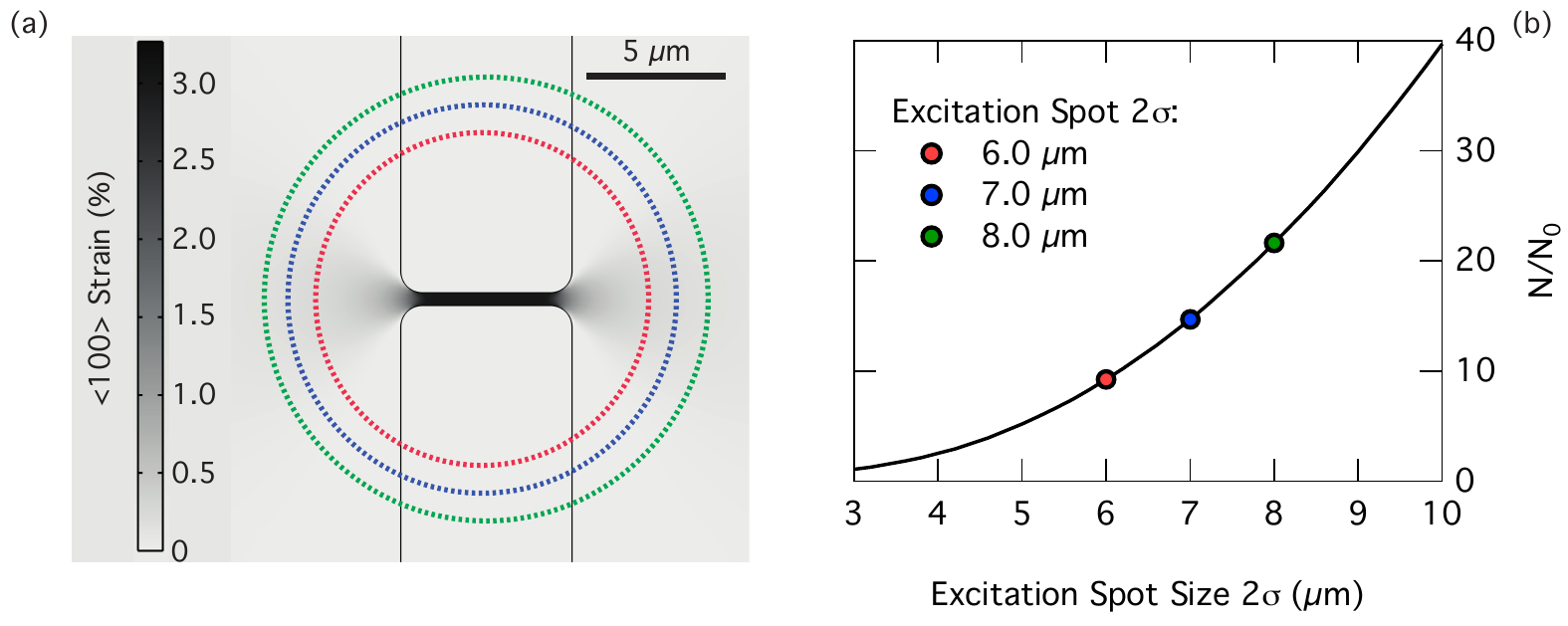}%
			\caption{(a) 2-dimensional $<$100$>$ strain map for a suspended GeOI microbridge. The circles in red, blue and green indicate the 1/e$^2$ limits of Gaussian excitation spots with 2$\sigma$ of 6.0 $\mu$m, 7.0 $\mu$m and 8.0 $\mu$m, respectively. (b) Increase in carrier density $N$ relative to carrier density $N_0$ in function of excitation spot size. The increase in carrier density stems from the diffusion of carriers into the strained constriction. The upper maximum of carrier increase is estimated by comparing the total excited area weighted with a symmetric 2-dimensional Gaussian distribution to the central 4.5 $\mu$m $\times$ 500 nm area with homogeneous strain.}
		\end{figure}

		\section{Strain Conversion from Raman Shift}
		
		Commonly, strain is converted with a linear relation from Raman-shifts. For the case of uniaxial stress along $<$100$>$, the strain $\varepsilon$ is related to the Raman shift $\Delta \omega$ as
		
		\begin{equation}
		\varepsilon = -\Delta \omega / 154 \ \mathrm{cm}^{-1} \label{lin_Raman}.
		\end{equation}
		
		However, recent experiments correlating Raman spectroscopy with strain via Laue microdiffraction show that for high uniaxial stress, there is a deviation from the linear relation [SM3]. \par
		As a comparison, the strain values of the investigated structures at 300 K and 20 K are given in table SMI for both, the linear and the non-linear conversion.
		
		\begin{table*}[h]
			\caption{Raman shifts measured for bridges aligned along $<$100$>$ together with corresponding strain values at 300 K and 20 K for conversion with the linear relation and non-linear relationship of Raman shift and strain.}
			\begin{ruledtabular}
				\begin{tabular}{c   c   c   c   c}
					$<$100$>$ Raman & $\varepsilon_\mathrm{300 K}$ (\%) & $\varepsilon_\mathrm{300 K}$ (\%)& $\varepsilon_\mathrm{20 K}$ (\%)& $\varepsilon_\mathrm{20 K}$ (\%) \\
					
					300 K & non-linear & linear & non-linear & linear \\
					
					$\Delta \omega$ $\left(\mathrm{cm}^{-1}\right)$ &conversion &  conversion& conversion& conversion \\
					
					\hline
					-3.3 & 1.9 & 2.1 & 2.8 & 3.2 \\
					-4.1 & 2.3 & 2.6 & 3.4 & 3.9 \\
					-4.9 & 2.7 & 3.2 & 4.0 & 4.7 \\
					-5.6 & 3.1 & 3.6 & 4.5 & 5.4 \\
					-6.7 & 3.6 & 4.3 & 5.4 & 6.4 \\
				\end{tabular}
			\end{ruledtabular}
		\end{table*}

	\begin{table*}
		\begin{tabular}{l l}
		$\left[\mathrm{SM1}\right]$ & \multirow{2}{17cm}{M. J. S{\"u}ess, R. Geiger, R. A. Minamisawa, G. Schiefler,
J. Frigerio, D. Chrastina, G. Isella, R. Spolenak, J. Faist,
and H. Sigg, Nature Photonics \textbf{7}, 466 (2013).}\\
  &  \\
		$\left[\mathrm{SM2}\right]$ & \multirow{2}{17cm}{D. Nam, D. S. Sukhdeo, J.-H. Kang, J. Petykiewicz, J. H. Lee, W. S. Jung, J. Vuckovic, M. L. Brongersma, and
K. C. Saraswat, Nano Letters \textbf{13}, 3118 (2013).}\\
		&  \\
		$\left[\mathrm{SM3}\right]$ & \multirow{2}{17cm}{A. Gassenq, S. Tardif, K. Guilloy, I. Duchemin, G. Osvaldo Dias, N. Pauc, D. Rouchon, J.-M. Hartmann, J. Widiez, J. Escalante, Y.-M. Niquet, R. Geiger, T. Zabel, H. Sigg, J. Faist, A. Chelnokov, F. Rieutord, V. Reboud, and V. Calvo, submitted.}\\
		&  \\  		  
	\end{tabular}
	\end{table*}

\end{document}